\documentclass[aps,pra,twocolumn,amsmath]{revtex4}
\usepackage{graphicx}
\newcommand{\beq}{\begin{equation}}
\newcommand{\eeq}{\end{equation}}
\newcommand{\bea}{\begin{eqnarray}}
\newcommand{\eea}{\end{eqnarray}}

\def\bfr{{\bf r}}

\def\hpsi{\hat \psi}
\def\hpsid{\hat \psi^\dagger}

\begin{document}

\draft \preprint{}
\title{Finite temperature theory of the trapped two dimensional
Bose gas}
\author{Christopher Gies$^1$, Brandon P. van Zyl$^2$, S. A. Morgan$^3$
and D. A. W. Hutchinson$^1$}
\address{$^1$Department of Physics, University of Otago,
P.O. Box 56, Dunedin, New Zealand
\\$^2$Department of Physics and Astronomy, McMaster University,
Hamilton, Ontario, Canada L8S 4M1
\\$^3$Department of Physics and Astronomy, University College London, Gower Street,
London WC1E 6BT, United Kingdom}
\date{\today}

\begin{abstract}

We present a Hartree-Fock-Bogoliubov (HFB) theoretical treatment of the
two-dimensional trapped Bose gas and indicate how semiclassical approximations
to this and other formalisms have lead to confusion. We numerically obtain
results for the quantum mechanical HFB theory within the Popov approximation
and show that the presence of the trap stabilizes the condensate against long
wavelength fluctuations. These results are used to show where phase
fluctuations lead to the formation of a quasicondensate.
\end{abstract} \pacs{03.75.Fi,05.30.Jp,67.40.Db}
\maketitle

The question of whether a weakly interacting Bose gas can undergo Bose-Einstein
condensation (BEC) when confined to an effectively two dimensional (2D)
geometry, has gained significant topical interest with the advent of recent
experiments \cite{gorlitz}. It is well know that a 2D  homogeneous ideal gas of
bosons does not undergo BEC at finite temperature \cite{statmech}. Indeed it
has been rigorously shown that even when interactions between the particles are
included \cite{hohenberg} there is no BEC phase transition at finite
temperature, although a superfluid phase transition in the form of a
Kosterlitz-Thouless (KT) transition \cite{KT} can be shown to take place. The
KT transition occurs because of the enhanced importance of fluctuations in the
2D system yielding a quasi-condensate state where the length scale for phase
coherence is small compared to the system size. The observation of such a phase
transition has been reported recently \cite{safanov} for a 2D gas of dilute
hydrogen on a liquid helium surface. For the non-interacting gas, Bagnato and
Kleppner \cite{Bagnato} showed that an ideal gas in a 2D harmonic trap does
exhibit a BEC phase. Petrov \emph{et al.$\,$}\cite{petrov} included
interactions and showed within the Thomas-Fermi approximation, that well below
$T_c$ there exists a true condensate while at higher temperatures a
quasicondensate forms. We are interested in the phase diagram at temperatures
below and within this fluctuation regime to investigate both the true BEC state
and the onset of phase fluctuations that lead to the destruction of the BEC.

 In three dimensions (3D), the finite temperature Hartree-Fock-Bogoliubov
(HFB) treatment of the trapped Bose gas has proved very successful
\cite{allan,rob,hzg}. It is therefore natural to develop this approach for the 2D
gas. Remarkably the quantum mechanical HFB formalism has, to the best of our
knowledge, never been implemented in 2D, the development of which is the central
result of this article. First let us consider however, a simplification to the fully
quantum mechanical implementation of HFB that has been used to conclude previously
that there is no BEC in an interacting 2D Bose gas.

In 2D, with a cylindrically symmetric trapping potential, the grand-canonical many-body
Hamiltonian is given by
\begin{multline}
H=\int \mathrm{d}^2r\; \hpsid(r)\\ \left(- \frac{\hbar^2 \nabla^{2}}{2m}+\frac{m
\omega_0^2}{2}\,r^2 - \mu+ \frac{g(r)}{2}\,\hpsid(r)\hpsi(r) \right ) \hpsi (r) ~,
\label{hamiltonian}
\end{multline}
where we have assumed a spatially dependent coupling parameter, introduced by the
many-body T-matrix for the trapped case as proposed in \cite{mark,sam}.

In a manner identical to the development of the theory in 3D \cite{gapless2} we
assume that the many-body Bose field operator can be decomposed into a mean,
condensate part and a fluctuating field operator part such that $\hpsi=\langle
\hpsi \rangle + \tilde \psi \equiv \phi + \tilde \psi$. The above Hamiltonian can
then be diagonalised provided the condensate order parameter, $\phi$, obeys the
generalised Gross-Pitaevskii equation (GPE) \beq \left ( - \frac{\hbar^2
\nabla^2}{2m} +\frac{m \omega_0^2}{2}\,r^2 - \mu +g(n_c+2 \tilde n) \right ) \phi
=0 ~, \label{gpe} \eeq and the quasiparticle excitations obey the Bogoliubov-de
Gennes (BdG) equations \bea \hat {\cal L} u_i-gn_cv_i&=& E_iu_i
\nonumber \\
\hat {\cal L} v_i-gn_cu_i&=& -E_iv_i ~, \label{BdG} \eea where $\hat {\cal L}= -
\frac{\hbar^2 \nabla^2}{2m}+\frac{m \omega_0^2}{2}\,r^2 - \mu +g(2n_c+2 \tilde
n)$, $n_c=\phi^*\phi$ is the condensate density and $\tilde n$ the noncondensate
density, which is evaluated by populating the quasiparticle levels according to
the usual Bose distribution. In the derivation of these equations we have used
the canonical transformation $\tilde \psi = \sum_i [u_i \hat \alpha_i - v_i^*
\hat \alpha_i^\dagger]$ where the $\hat \alpha_i$ satisfy the usual Bose
commutation relations and we have taken the so called Popov approximation,
neglecting anomalous pair averages of the fluctuating field operator.

Together, the GPE and BdG equations form a closed set and can be solved numerically
using techniques analogous to those developed in 3D. This we will describe shortly,
but first let us attempt to obtain a semiclassical solution of these equations. This
is the approach taken by previous authors \cite{mullin}.

Let us assume that the temperature is large compared to the energy level spacing in
the trap. We can now replace the quasiparticle amplitudes such that $u_i \approx
ue^{i \theta}$ and $v_i \approx ve^{i \theta}$ where the common phase, $\theta$,
defines a quasiparticle momentum, ${\bf p} = \hbar \nabla \theta$. Neglecting
spatial derivatives of $u$, $v$, and $\bf{p}$ (local density approximation) and
making a continuum approximation for the Bose distribution function, the
quasiparticle excitation spectrum is found to be given by \beq E_{sc}({\bf p},
\bfr)= \sqrt{\Lambda^2_{sc}-(gn_c)^2} ~, \label{scbog} \eeq where
$\Lambda_{sc}=\frac{{\bf p}^2}{2m}+\frac{1}{2}m\omega_0^2r^2-\mu+2gn_c+2g \tilde n$
and the condensate density is still calculated via the GPE. The density of excited
state particles can now be integrated in closed form to give \beq \tilde n =
\frac{1}{\lambda^2} \left [ -\ln \left (1-\exp(-\sqrt{t^2-s^2}) \right )
+\frac{t}{2}- \frac{1}{2}\sqrt{t^2-s^2} \right ] ~, \label{scnoncon} \eeq where
$\lambda^2=\hbar^2/2\pi m k T$, $s=gn_c/kT$ and $t=\frac{1}{kT}\left ( \frac{1}{2}m
\omega_0^2r^2-\mu+2gn_c\right )$. These semiclassical HFB equations form a closed
set and it is this set of equations that previous authors have attempted to solve
self-consistently. This is not possible however, as the semiclassical approximation
can only be used consistently at low energy if combined with a Thomas-Fermi
approximation for the condensate \cite{reidl}. There are therefore no solutions to
these equations.

It is trivial to show that in the Thomas-Fermi limit $t=s$ and the expression for
the semiclassical thermal density is undefined. Indeed, even for low particle
numbers, it can be shown that the arguments of the square roots always contain
values at some spatial point that are negative (and approach zero from below in the
Thomas-Fermi limit) and hence this expression is never well defined. The origin of
this problem lies in the expression for the semiclassical excitation spectrum. At
low energies, or equivalently long wavelengths, Eq.\ (4) yields imaginary energies.
This has been used to conclude that in 2D BEC cannot take place since the condensate
is destabilised by long wavelength fluctuations \cite{mullin}. This is simply
incorrect. What one is seeing is a failure of the semiclassical approximation. If
the discrete nature of the excitation spectrum for the finite size trap is not
retained for the low energy excitations, and this is the case within the
semiclassical approximation, there comes a point where the magnitude of the chemical
potential exceeds the magnitude of the effective potential and the argument in Eq.\
(4) becomes negative for low {\bf p}. Just because the semiclassical treatment of
the HFB formalism fails does not mean that the quantum mechanical theory will also.
In this case the long wavelength oscillations, corresponding to the ${\bf p}={\bf
0}$ limit in the semiclassical case, are precluded by the finite size of the trap.
Therefore, to determine whether BEC can take place in 2D, the full, numerically
expensive, discrete calculation must be undertaken.

As an aside, it is possible to obtain a well defined semiclassical theory of
the trapped 2D gas if one makes the further Hartree-Fock approximation
\cite{kim}. This consists of setting the $v(\bfr, {\bf p})$ terms in the
semiclassical HFB treatment to zero everywhere. The Hartree-Fock excitation
spectrum is now single particle-like and is just given by $\Lambda_{sc}$ rather
than by the phonon-like Bogoliubov spectrum of Eq.\ (4). There is therefore no
problem with the infrared divergence seen previously and one can integrate to
obtain the semiclassical Hartree-Fock thermal density, \beq \tilde n= -
\frac{1}{\lambda^2}\left [ \ln \left (1-e^{-( \frac{1}{2}m\omega_0^2r^2
+2gn_c+2g \tilde n -\mu)/kT} \right ) \right ] ~, \label{hfscnoncon} \eeq which
can be solved self-consistently together with the GPE to obtain a Bose
condensed solution. Interestingly, if one omits the ground state from the
calculation, effectively demanding that condensation {\it does not} occur, then
it is still possible to obtain a self-consistent solution for the thermal
density at all temperatures. Therefore, at the level of the semiclassical
Hartree-Fock approximation, one might conclude that there is no BEC in the
thermodynamic limit for a 2D gas of trapped, interacting bosons.

By removing the lowest lying state however, one has truncated the available state
space and so one may expect that the thermodynamically stable state is the condensed
phase. This is confirmed by calculating the free energies for the two states. Above
the BEC transition temperature the free energy of the condensed and uncondensed
solutions are virtually indistinguishable, but below the critical temperature the
free energy for the condensed solution is the lower, confirming that within the
Hartree-Fock approximation, BEC is the thermodynamically favoured state
\cite{mullin}.

We now proceed to solve the quantum mechanical, discrete, HFB equations [Eq.\
(2) and Eq.\ (3)] self-consistently. The method of solution is described in detail
elsewhere \cite{gapless2}, but is outlined here for completeness.

First, the GPE equation is solved using an expansion in some appropriate basis set
with the condensate and noncondensate densities set to zero. The solution for the
condensate wavefunction is then used to construct $n_c$ and the process iterated to
find a converged solution. One now needs to calculate the noncondensate density. To
do this one decouples the BdG equations by making the transformation to the
auxillary functions $\psi_i^{(\pm)} \equiv u_i \pm v_i$. One can thus obtain
equations for $\psi_i^{(+)}$ and $\psi_i^{(-)}$ separately. These are solved using a
further basis set expansion, for which we use the basis of excited states of the GPE
to ensure orthogonality with the condensate and to simplify the construction of the
matrix elements. The noncondensate density is then constructed by populating the
quasiparticle states. This value of $\tilde n$ is inserted into the GPE and the
whole process repeated iteratively until convergence.

In contrast to the semiclassical HFB treatment we have no difficulty in finding
self-consistent solutions to the quantum, discrete HFB equations and typical
results are presented below. In the numerical solution we non-dimensionalise our
equations using the natural harmonic oscillator units. In this case we take our
Rydberg of energy to be $\hbar \omega_0/2$ and in these units the non-dimensional
interaction parameter, $g(r)$, takes values between $0.09$ and $0.1$. This
parameter, in 3D, depends only upon the s-wave scattering length, $a$, however, in
2D, $g$ also depends upon the strength of the confining potential in the third
dimension. The 2D gas thus represents a system where the interparticle interaction
strength can be tuned by modulating the confining potential in the z-direction in a
manner analogous to the use of the Feshbach resonance in 3D \cite{ketterle}. It has
even been suggested that the sign of the interaction strength (and hence whether the
interactions are attractive or repulsive) can be changed by tuning the confining
potential in the third direction \cite{petrov}.
%
\begin{figure}
\begin{center}
\vspace{-.5cm}
\includegraphics[width=\columnwidth,height=6.2cm]{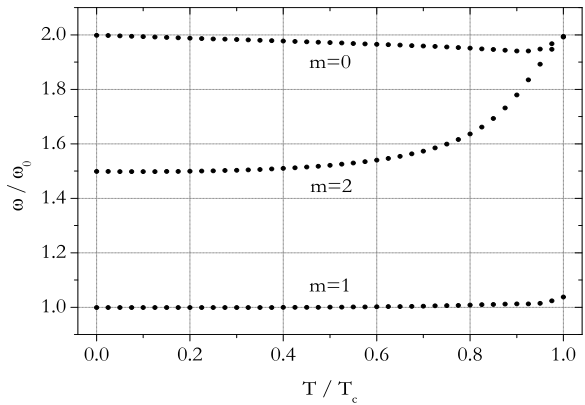}
\end{center}
\vspace{-1cm} \caption{Quasiparticle excitation frequencies as a function of
temperature for $N=2000$ atoms and the full spatially dependent coupling
parameter.} \vspace{-3mm}
\end{figure}

In Fig.\ (1) we display the low lying excitation spectrum as a function of
temperature for the trapped gas of 2000 $^{87}$Rb atoms. Shown are the lowest
lying $m=1$, $m=2$ and $m=0$ quasiparticle, or collective, modes up to the
critical temperature. We point out that the results shown are those of the
HFB-Popov equations. In three dimensions, in the presence of significant direct
driving of the thermal cloud, the theory fails to predict shifts in the
frequencies above about $0.6\,T_c$ \cite{sam_2ndorder}. These are related to the
dynamics of the thermal cloud not included in this theory. The excitation
spectrum shown should be valid in experiments where the condensate is driven
directly.

In the three dimensional asymmetric case this approximation is valid only below
$0.6\,T_c$. However, we include the higher temperatures for the symmetrical
case where the theory in 3D does not fail.
The $m=1$ mode is the Kohn mode \cite{kohn} and clearly satisfies the generalised
Kohn theorem to within our resolution, apart from a slight deviation near the
critical temperature. The quadrupole ($m=2$) and breathing ($m=0$) modes are well
defined and nowhere look like going soft (approaching zero frequency as an
indication of an instability). We therefore conclude there are well defined
solutions to the quantum mechanical HFB equations, in a manner exactly the same as
in 3D, for all temperatures below the critical temperature.

It is important to note that, at low temperatures, the frequency of the lowest lying
$m=0$ mode, or breathing mode, is at precisely $2 \omega_0$, independent of the
interaction strength. This result was predicted some time ago by Pitaevskii and
Rosch \cite{pitaevskii}, purely from symmetry arguments. Indeed, they show
rigorously that the existence of $2 \omega_0$ oscillations is ensured by the
underlying SO(2,1) symmetry of the {\it full quantum theory Hamiltonian} for the
interacting harmonically confined 2D gas with a contact interaction. It is therefore
noteworthy (and critical) that our calculations confirm this result numerically at
low temperatures. At higher temperatures the result is still valid, however the
excitations are those for a condensate in an effective potential which is modified
by the addition of the potential $2g \tilde n$ from the static thermal cloud. The
condensate effectively sees a weaker harmonic potential and hence the $m=0$ mode has
a slightly lower frequency. Above the critical temperature, the excitation
frequencies, of course, go over to those of the thermal gas. Similarly, at high
temperatures (near the critical temperature) the $m=1$ mode no longer satisfies the
Kohn theorem precisely. This is again due to the presence of the potential from the
thermal cloud. The effective potential in which the condensate oscillates is no
longer harmonic and the Kohn theorem broken. As discussed previously
\cite{gapless2}, if the full dynamics of the thermal cloud were included, then the
Kohn theorem would be satisfied at all temperatures.


\begin{figure}
\begin{center}
\vspace{-1cm}
\includegraphics[width= \columnwidth]{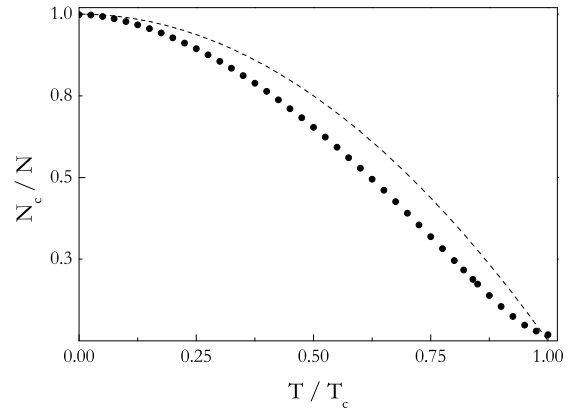}
\end{center}
\vspace{-1cm} \caption{Fraction of atoms in the condensate as a function of
temperature. The dotted line corresponds to the non-interacting gas for
comparison.}\label{fig:frac} \vspace{-4mm}
\end{figure}

\begin{figure}
\begin{center}
\vspace{-.8cm}
\includegraphics[width=\columnwidth]{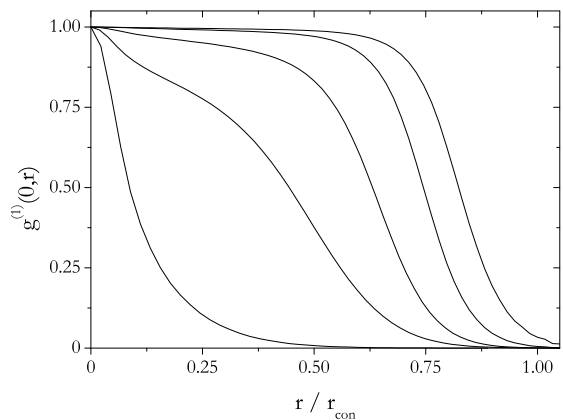}
\end{center}
\vspace{-1cm} \caption{Single particle correlation function for the 2D Bose gas as a
function of position at 0.05, 0.1, 0.35, 0.75 and 0.925 $T_c$ from right to left, showing
decreasing correlation length as a function of temperature. Lengths are scaled with the
size of the condensate, $r_\mathrm{con}$,
 at each temperature.}\label{fig:g1}
 \vspace{-2mm}
\end{figure}

In Fig.\ \ref{fig:frac} we show the condensate fraction as a function of temperature. It
is clear that at temperatures below the critical temperature we obtain a macroscopic
occupation of the ground state, which implies BEC. This however is not sufficient. For
true condensation we require a well defined phase over the entire condensate. In 3D this
is true for all temperatures below $T_c$ except for a small region near the critical
temperature often refered to as the Ginzburg region. In this region phase fluctuations
prohibit the formation of a true condensate. In a uniform 2D gas this region extends all
the way to $T=0$ and is what prevents the formation of a BEC. In Fig.\ \ref{fig:g1} we
plot the off-diagonal correlation function, $g^{(1)}(0,r)$ \cite{nara} showing that only
at low temperatures is there a {\em coherent} condensate with a correlated phase spanning
the condensate. As the temperature is increased the coherence begins to decay on a length
scale less than the dimensions of the gas. In analogy with Petrov {\it et al.}
\cite{petrov}, if we expand the field operator as $\hat{\psi} =
\phi_0+\delta\hat{\psi}=\sqrt{\hat{n}}e^{i\hat{\phi}}$ then we can express the
noncondensate density as $\tilde{n} = \langle
\delta\hat{\psi}^{\dagger}\delta\hat{\psi}\rangle = \langle \frac{\delta\hat{n}^2}{4n_0}
+ i[\delta\hat{n},\hat{\phi}]/2 + n_0\hat{\phi}^2 \rangle$. At low temperatures density
fluctuations are suppressed and this yields $\tilde{n}/n_0 \sim \langle \hat{\phi}^2
\rangle$. Using the Thomas-Fermi approximation  Petrov {\it et al.} conclude that phase
fluctuations lead to the formation of a quasicondensate at temperatures of around
$T=T_c/2$ for our parameters. In our case we find that the condensate persists to
approximately this temperature, but above this the phase becomes ill defined and the
system is best described as a quasicondensate. This is consistent with our numerical
calculation of the correlation function. A treatment of the 2D gas which could describe
the quasicondensate has recently been formulated \cite{nick}, but has yet to be
implemented.

In conclusion, we have shown that the presence of the trap stabilizes the condensate
against long wavelength fluctuations. This is true not only for density fluctuations, but
for phase fluctuations as well, which are included in our formalism via the contribution
to the non-condensate density from low-energy quasiparticles. Our work is consistent with
Petrov {\it et al.\ }\cite{petrov}. A 2D trapped dilute gas of weakly interacting bosons
therefore does undergo BEC, forming a pure condensate at temperatures below a transition
region near $T_c$. Although this conclusion has been reached by, among others, Petrov
{\it et al.\ }\cite{petrov} and Bagnato and Kleppner \cite{Bagnato}, the converse
conclusion has also appeared in the literature \cite{mullin}. The prospect of performing
the HFB calculation has been proposed as a means of clarifying the issue (in addition to
the above references, see for example, Bayindir and Tanatar \cite{Bayindir}), but to this
point no one had done so. We have now performed this calculation and unambiguously shown
that BEC does occur for the 2D trapped interacting gas when the discrete nature of the
energy spectrum is taken into consideration.

This work was funded through a University of Otago Research Grant and by the Marsden
Fund of the Royal Society of New Zealand. BvZ would like to acknowledge support from
Rajat Bhaduri through the NSERC of Canada. SM thanks the Royal Society (London) for
financial support and CG the DAAD for partial support. DH would like to thank
Matthew Davis for many stimulating and useful discussions.

\end{document}